\documentclass[twocolumn,showpacs,showkeys,preprintnumbers,amsmath,amssymb]{revtex4}

\usepackage{graphicx}
\usepackage{dcolumn}
\usepackage{bm}

\begin{document}


\title{Dynamical formation of a nonequilibrium subsystem under severe action}

\author{Leonid S.Metlov}
\email{lsmet@kinetic.ac.donetsk.ua}
\affiliation{Donetsk Institute of Physics and Engineering, Ukrainian
Academy of Sciences,
\\83114, R.Luxemburg str. 72, Donetsk, Ukraine}

\thanks{The work was supported by the budget topic № 0106U006931 of NAS of Ukraine and partially by the Ukrainian state fund of fundamental researches (grants F28.443-2009). The author thanks A. Filippov for helpful discussions.}

\date{\today}

\begin{abstract}
Formation of the nonequilibrium subsystem in dynamical processes during defect generation is simulated by means of molecular dynamics. A particular process of dissipation of the low-frequency acoustic emission into high-frequency equilibrium vibrations of lattice is studied numerically. Clear heuristic reasons are used to introduce different temperatures and entropies for equilibrium and nonequilibrium subsystems. Simple relaxation equation is proposed to describe time behavior of the nonequilibrium entropy.
\end{abstract}


\maketitle

\section{Introduction}

Well known problem of the statistical physics and thermodynamics \cite{tsal00,ps84} is related to reducing of the reversible laws of the micromechanics into irreversible laws of nonequilibrium thermodynamics \cite{ee59,kryl79,dfku04}. It invokes some simplifying assumptions which solve different particular problems, but leads to a loss of generality. After many years of the study it still remains reasonable to seek for the new ideas and, maybe, return to initial sources of the problem. In solids, for example, with their complex internal structure the problem is related to existence of weakly interacting subsystems. In many cases one can define relatively small spatially separated subsystems where an equilibrium state establishes quicker than in the whole system (so called, local thermodynamic equilibrium) \cite{gp71,zmr02}. In another cases one can define spatially overlapped subsystems, which have their own temperature and other thermodynamic parameters. For example, it takes place for the electron and phonon subsystems, which can be almost totally separated in the adiabatic approach \cite{cw71}. The same can happen for the phonon subsystem and radiation \cite{w01}, or in particular for acoustic emission \cite{gt05}. In all these cases every subsystem may be in the equilibrium state, but complete system is not equilibrated.

Formation of the defects in the solids under external impact provokes nonequilibrium processes. It transfers a part of the internal energy into thermal form. However, this part does not transform into heat immediately and radiates in the first stages in a form of low-frequency acoustic waves. These waves gradually dissipate and supply the energy to equilibrium thermal motion. As result, one can distinguish two weakly interacting subsystems of high-frequency phonons (vibration modes) and the low-frequency acoustic emission. First subsystem is normally equilibrium before the external impact, and remains almost equilibrium after it. The second one is created by the impact and strongly nonequilibrium. Present article is devoted to study of the interaction between these two equilibrium and nonequilibrium subsystems by means molecular dynamics simulation.

\section{MD-simulation}

\begin{figure}[h]
  \center{\includegraphics [width=3.0 in] {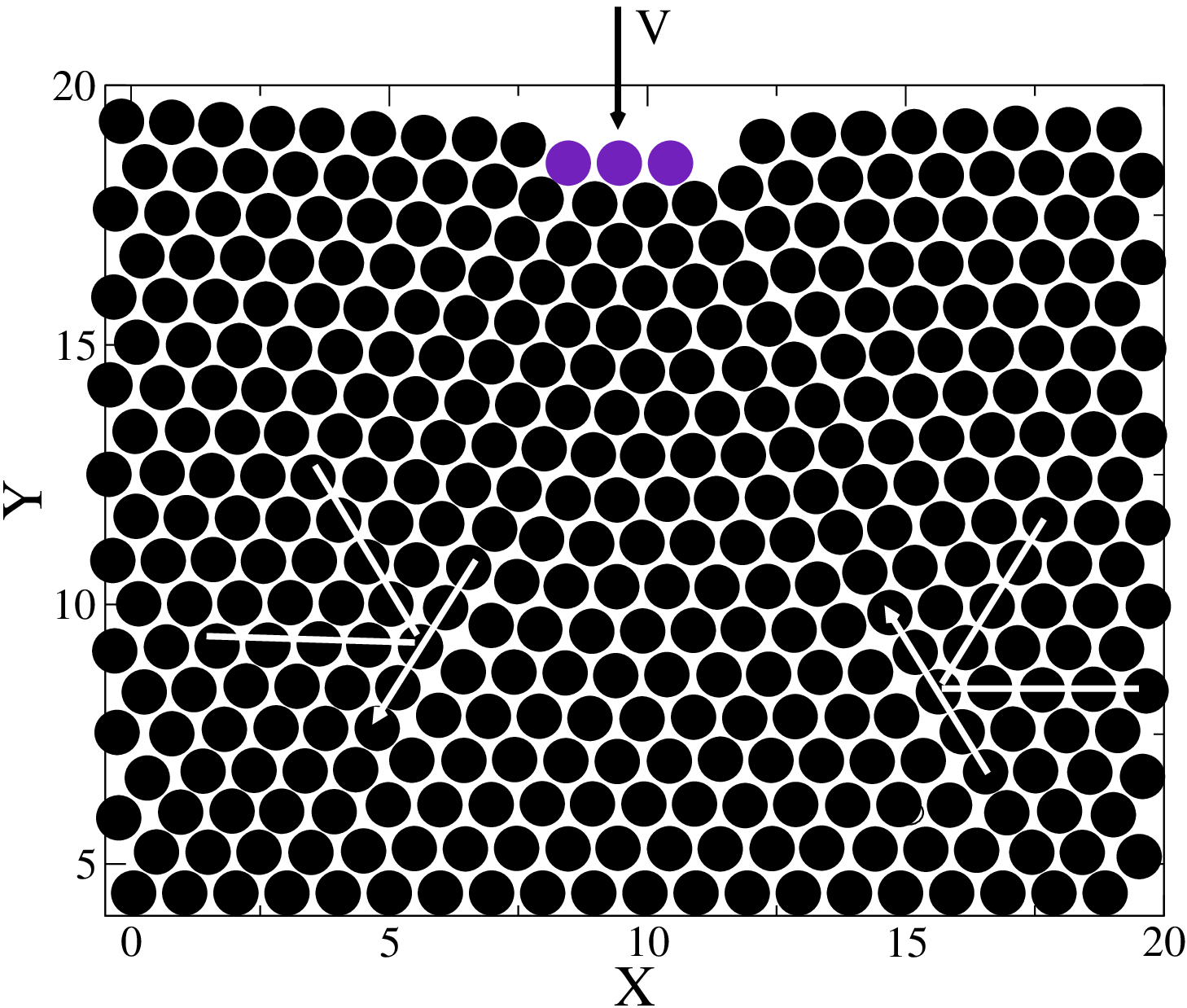}}
  \caption{Geometry of 3D computer experiment – positions of atoms at the 2900th time step. Time step is equal to 10.6 Fs.}
  \label{f1}
\end{figure}

\begin{figure*}
  \includegraphics [width=5.0 in] {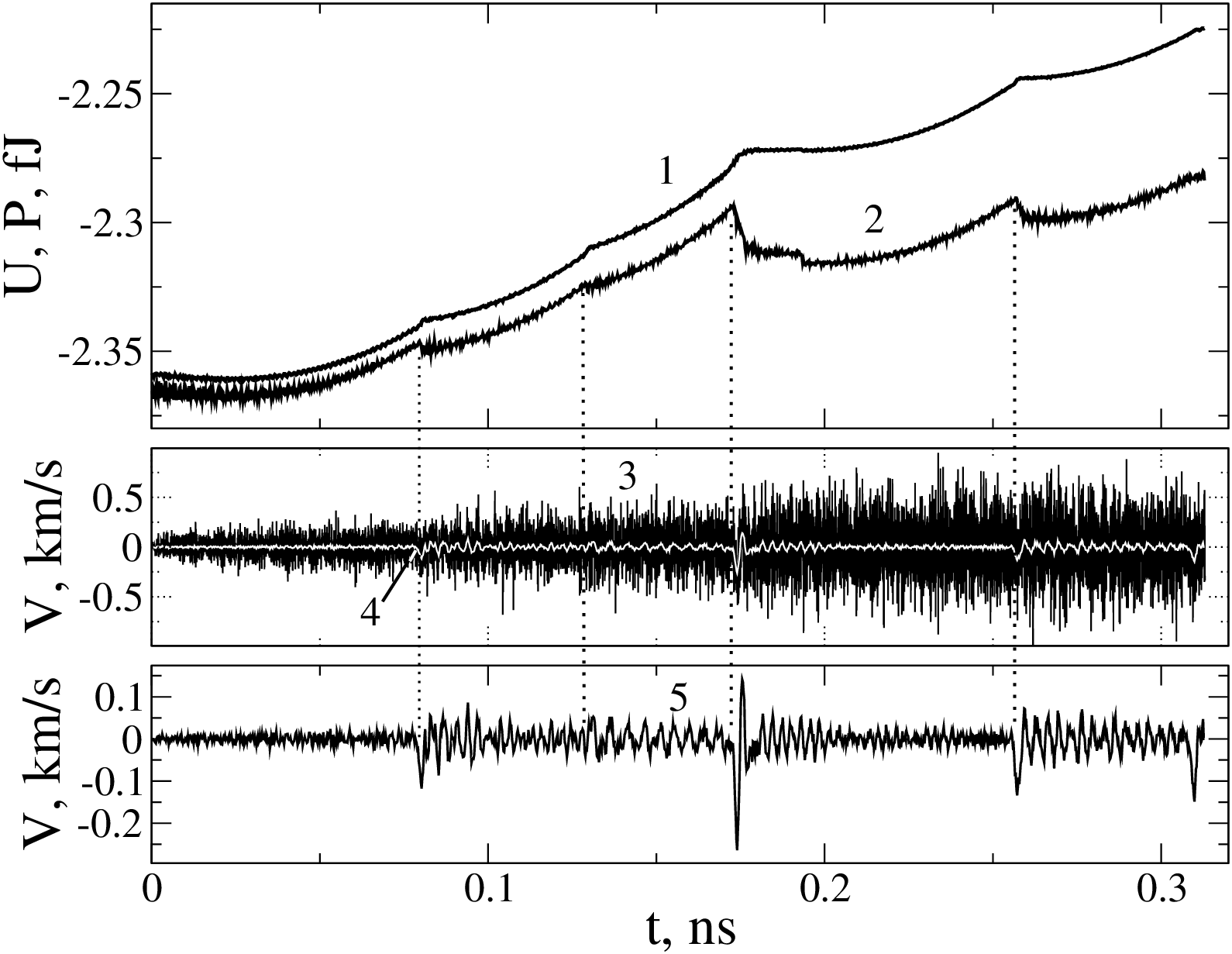}
  \caption{Time scanning: 1 – total (internal) energy; 2 – total potential energy; 3 – velocity of arbitrary particle; 4, 5 – low-frequency record of particle velocity after filtration. 
}
\label{f2}
\end{figure*}

\begin{figure*}
  \includegraphics [width=3.0 in] {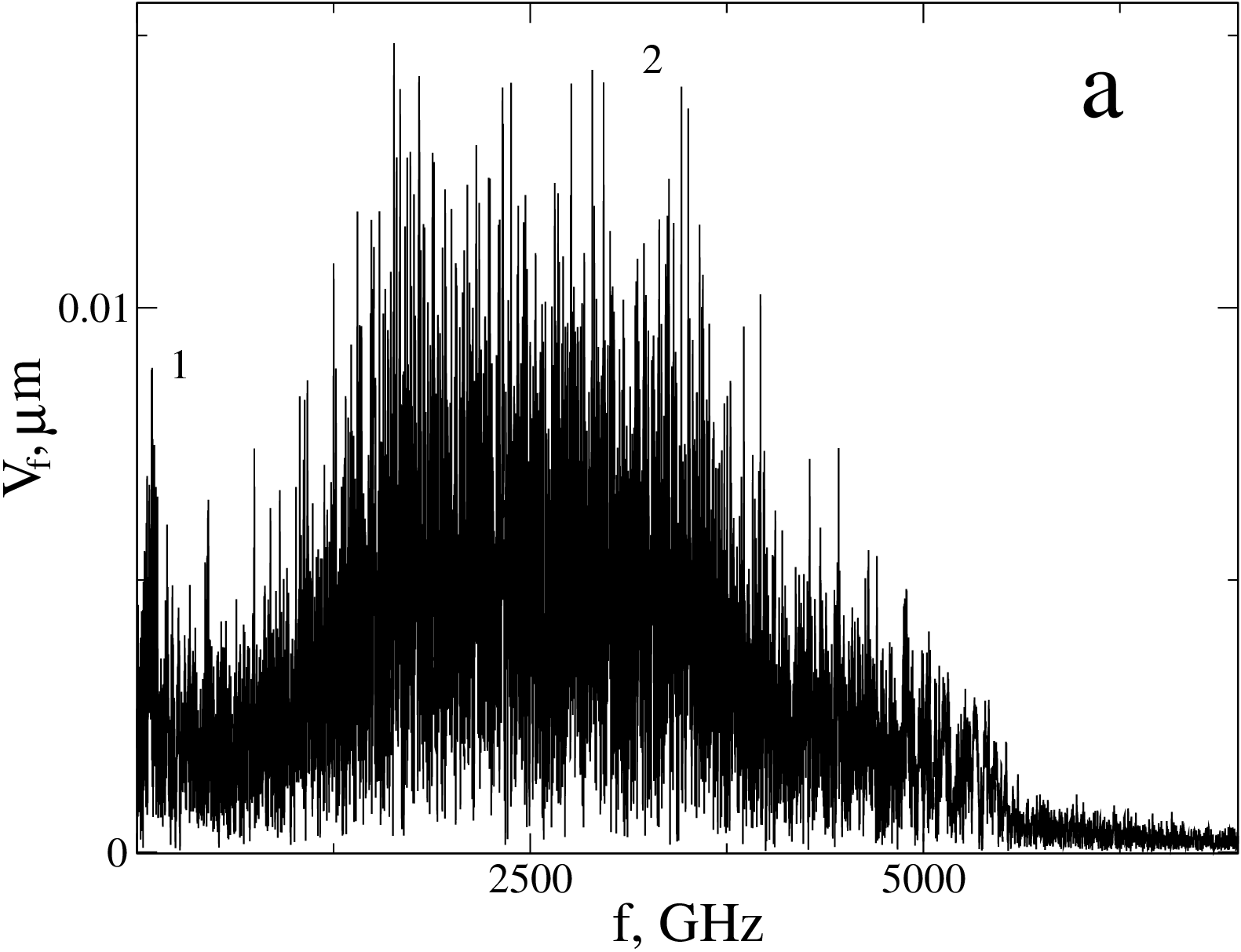}
  \includegraphics [width=3.0 in] {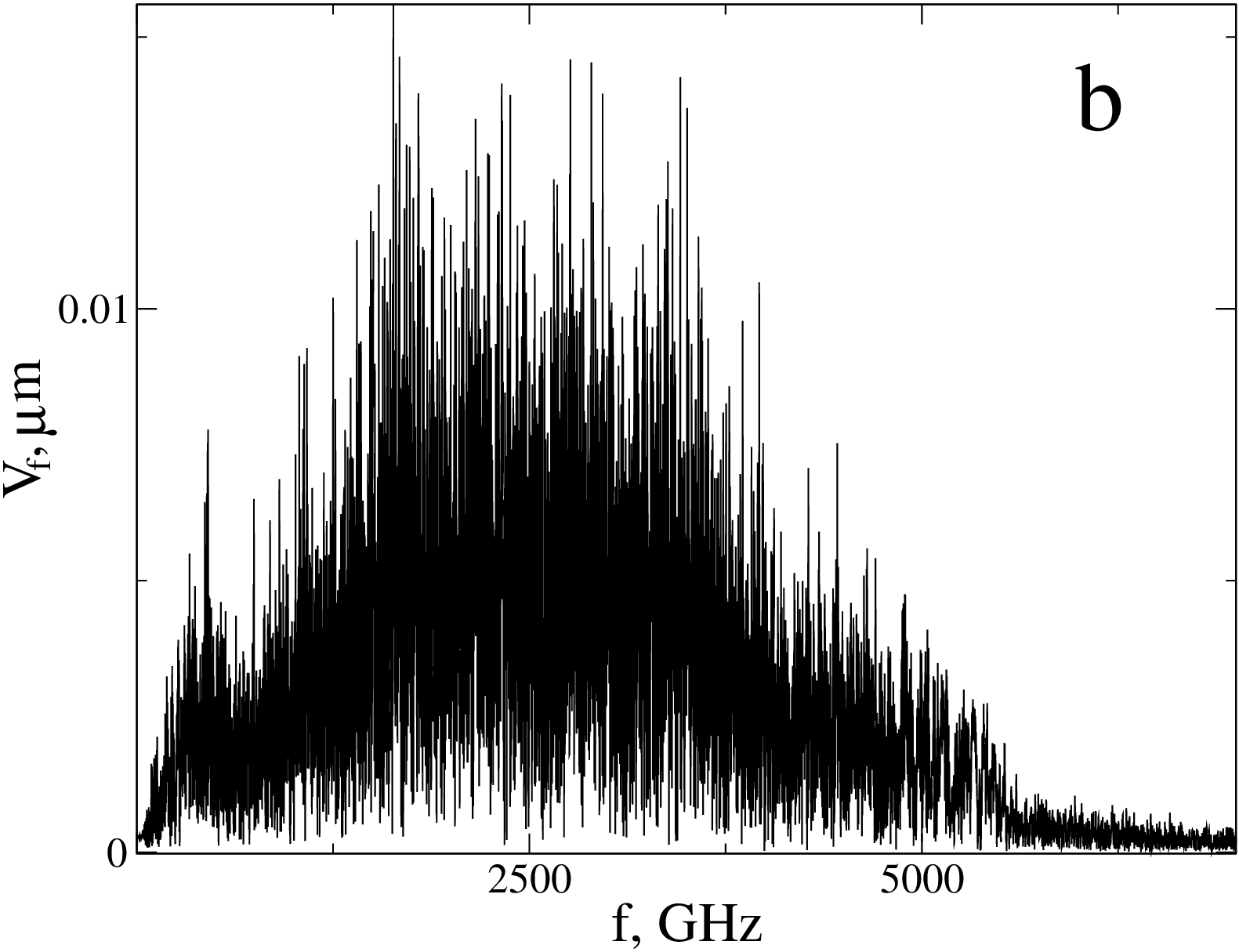}
  \caption{Spectrum of total record (a) and difference one (b): 1- law frequency peak; 2 - main region of spectrum
}
\label{f3}
\end{figure*}

Let us regard to a particular configuration of the hard indenter pressing into 3D atomic specimen (Fig. \ref{f1}). The indenter (three upper atoms in the planar projection) moves with constant velocity 5m/s. The atoms possessed by lowest array are motionless.  Behind the surface shown in the front of the picture 3D system contains 10 closed packed planes of atoms placed into HCP structure. All lateral faces are free. The sample consists from copper atoms interacting by Lennard- Jones potential. At some time moment the dislocations emerge in the atomic specimen.  Fig. \ref{f1}
presents exactly the stage when the dislocations start to enter into the system. It manifests itself by a bend of the curve for time depending total internal energy (curve 1, Fig. \ref{f2})
or in the jumps of total potential energy (curve 2). Second subplot presents time dependence of the velocity for arbitrarily chosen particle of the system (which actually records its thermal motion in curve 3). The thermal motion can be treated as high-frequency signal which is close to a harmonic one for separate interatomic (bond) vibrations. Its spectrum is severely oscillated in consequence of interference of large set of high-frequency phonons (region 2, Fig. \ref{f3}a).
In the low-frequency region a large peak stands up sharply against the rest spectrum (peak 1). Performing low-frequency filtration over 100 time steps one can separate the low-frequency vibration modes (curves 4 and 5 in the last subplot). These modes gradually dissipate with time. Comparing curves 1-5 one can see directly that initiation of the low-frequency vibrational excitations coincides in time with a generation of the dislocations and can be associated in this sense with strongly nonlinear acoustic emission.

The external work performed during indentation increases total internal and potential energies (beginning parts of curves 1 and 2). Further, after t = 0.075ns, the potential energy decreases sharply (curve 2). A part of this energy transforms into defect energy and results in dislocation nucleation. Remaining part dissipates into heat motion via radiation in a form of the low-frequency vibration waves (curves 4 and 5 respectively). These acoustic waves scatter on high-frequency phonons and as result are damped exponentially. This process of the nonequilibrium state relaxation gradually leads to slow increase of equilibrium temperature and entropy.

During indentation the entrance of the dislocations repeats from time to time (see points t = 0.125ns; 0.175ns; 0.26ns in Fig. \ref{f2}). Due to this some fraction of the low-frequency nonequilibrium wave packets always presents in integral thermal motion. Low-frequency component of wave motion differs from high-frequency phonon field by spatial and time scales only. So, to describe it one can use the same language as for the ordinary phonons. In particular, thermal motion of the phonons is characterized by temperature and entropy. The temperature can be treated in both manners: as mean energy per one degree of freedom (averaged over ensemble) or as mean energy of the vibrations of each particle (averaged over time) \cite{c06,c08}. To get equilibrium velocity of the particles one can subtract a record of nonequilibrium low-frequency vibrations (curve 4, Fig. \ref{f2}) from the complete one (curve 3).

\begin{figure}
  \includegraphics [width=3.0 in] {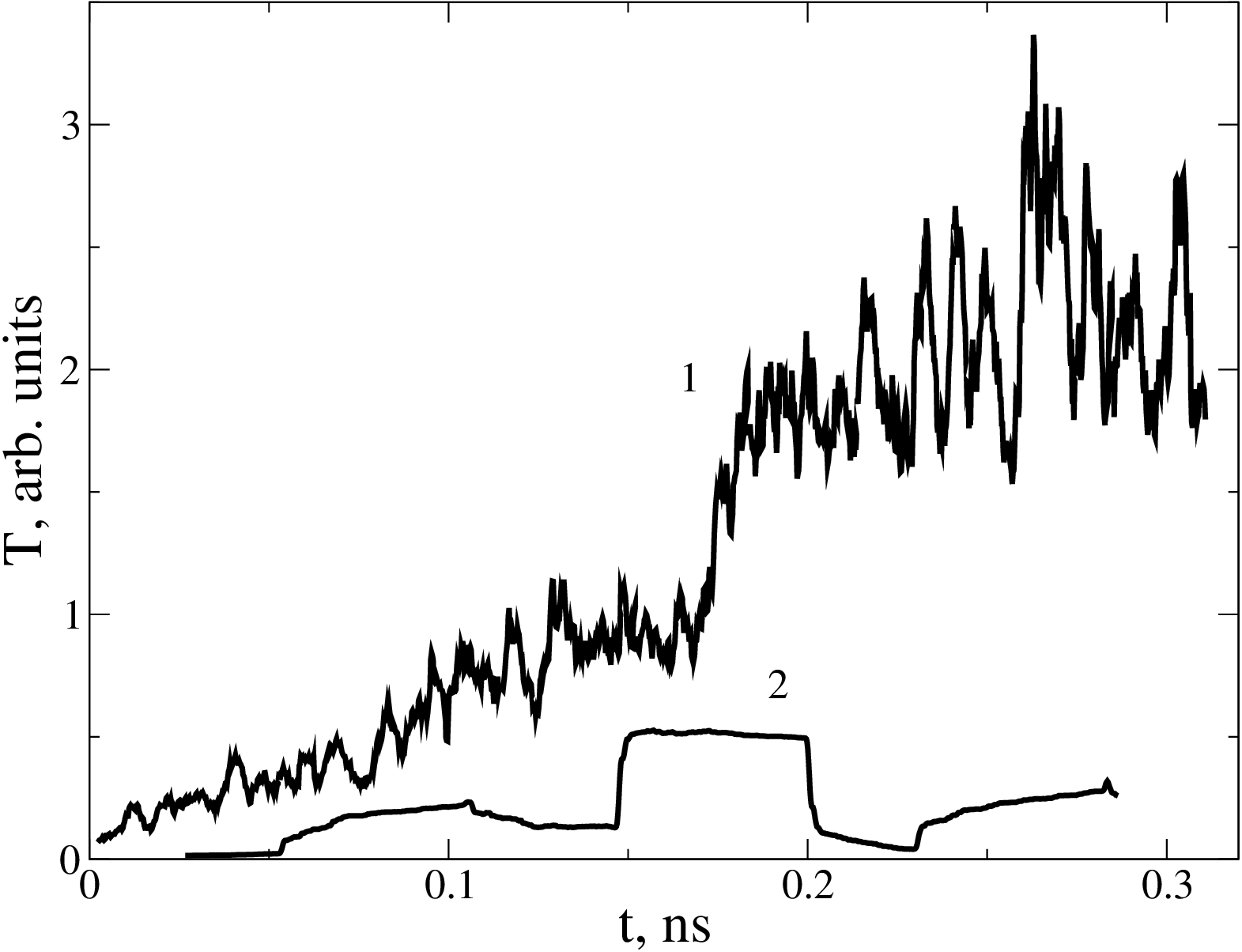}
  \caption{Time dependence: 1 - equilibrium temperature; 2 - nonequilibrium temperature}
\label{f4}
\end{figure}

To control the procedure let us recalculate remaining energy the spectrum. It is shown in Fig. \ref{f3}b), where the peak 1 corresponding low-frequency vibration packets vanishes. Such a curve in Fig. \ref{f4}
 is plotted after average over interval of 100 time steps contains about 20 periods of the high-frequency vibrations, and seems to be enough to smooth of random fluctuations of the thermal motion.  One can notice from the figure that equilibrium temperature increases as energy of nonequilibrium low-frequency vibration transfers into equilibrium high-frequency motion. As low-frequency wave or vibration motion differs from high-frequency one by time scale only, one can use this observation to define a nonequilibrium temperature. For a large enough numerical system (and certainly for the real solids) it is possible to take so long enough time average to suppress completely random fluctuations belonging to the short range scales and separate large-scale perturbations caused by the entrance of the dislocations. Thus, digital filtration allows us separate equilibrium and nonequilibrium components from the initial data.

\section{Phenomenological model}

Let us now apply phenomenological approach to define nonequilibrium entropy. Kinetic equation for production and annihilation entropy   in nonequilibrium subsystem may be written as follows \cite{m08,m07}:
\begin{equation}\label{a1}
\dfrac{\partial \tilde{s}}{\partial t}=\gamma_{\tilde{s}} (T_{0}-T_{st}-T_{1}\tilde{s}),
\end{equation}
Here $T_{0}$ is temperature of internal sources producing nonequilibrium heat at, so $\gamma_{\tilde{s}}T_{0}$  is an intensity of the entropy sources for the nonequilibrium subsystem, $T_{st}$  is temperature of internal acceptors of the dissipation for nonequilibrium heat and $\gamma_{\tilde{s}}T_{st}$  is power of entropy sinks for the nonequilibrium subsystem, $T_{1}\tilde{s}$ is relaxation term. The value $T_{st}$  has stationary meaning of nonequilibrium temperature to which the system tends during evolution. The last term in Eq.\ref{a1} describes transfer of the entropy from nonequilibrium subsystem to equilibrium one due scattering of low-frequency vibrations on high-frequency phonons.

The validity of the above assumptions has been checked using  another example of acoustical emission at formation of corrosion microcracks into bulk of welded seam. As a result of numerical simulations the distribution of number of acoustic emission signals n on amplitude A during a kinetic stage is found to be described by Gaussian distribution \cite{vbml05}:
\begin{equation}\label{a2}
n(A)=n_{0}exp\left( -\frac{\left( A-\langle A \rangle \right)^2}{2\sigma^2} \right),
\end{equation}
where $n_{0}$ is a normalization, $\langle ...  \rangle$  is averaging operator, $\sigma$ is a standard deviation. The square of signal amplitude is energetic parameter per one act of acoustic emission. So, the relation (2) coincides with standard Gibbs distribution which standard deviation has a meaning of the “temperature” $T=2\sigma^2/k_{B}$.  In a stationary state one can assume $T=T_{st}$, where $T_{st}$ is stationary temperature involved into kinetic equation (\ref{a1}).

The same relations have been recently restored from another example of acoustic emission appearing during high-temperature deformation of aluminum. The acoustical emission is caused here by Porteven-le Shatelie effect related to a short-run coherent slip of large family of dislocations which form single slip trace and to a spatial correlation between deformation processes in mesoscopic scale \cite{bl93,lb98}. Temperature   dependence of activation volume   the has been derived in the Ref. \cite{pm08} in the form:
\begin{equation}\label{a3}
y=y_{0}+Aexp\left( \dfrac{T}{t} \right),
\end{equation}
where $y_{0}=0.47 \pm 0,12$,  $A=0.0027 \pm 0,0004$,  $t=130 \pm 39$. If the temperature is high enough one can neglect additive term $y_{0}$, expand the exponential over temperature  $T_{r}$ and reduce the dependence (\ref{a3}) to the effective Gibbs distribution:
\begin{equation}\label{a4}
n=n_{0}exp\left( -\dfrac{\varepsilon}{k_{B}T} \right),
\end{equation}
Here $n$  is number of separated pulses of acoustical emission, forming given region of activation volume, $\varepsilon$  is activation energy of acoustic emission and the constants in the equations (\ref{a3}) and (\ref{a4}) are related as follows:
\begin{equation}\label{a5}
    A=n_{0}exp \left( -\dfrac{\epsilon}{k_{B}T_{r}} \right) ,
    \quad t= \dfrac{k_{B}T_{r}^{2}}{\varepsilon},
\end{equation}
Thus, constant $t$ has a meaning of inverse activation energy.

To summary: using molecular dynamic simulation we have investigated a mechanism of nonequilibrium subsystem formation due to dislocation defect generation. This process produces low-frequency nonequilibrium phonons (acoustic emission) which dissipate into high-frequency equilibrium lattice vibrations. The temperatures of equilibrium and nonequilibrium subsystems (as well as conjugated to them equilibrium and nonequilibrium entropies) have been introduced on a base of clear heuristic reasons and illustrated using some additional examples of the acoustic emission. Simple relaxation equation is proposed to describe the nonequilibrium entropy evolution. A comparison of the results with experimental studies of acoustical emission on another structural scale is performed.


\begin{thebibliography}{00}
\bibitem{tsal00} C. Tsallis. Nonextensive Statistical Mechanics and Thermodynamics: Historical Background and Present Status, in the book: «Lecture Notes in Physics»,  Springer, Berlin, Heidelberg, New York, Barcelona, Hong Kong, London, Milan, Paris, Singapore, Tokyo, 2000.
\bibitem{ps84} I. Prigogine, I. Stengers, Order out of Chaos \& Men's New Dialogue with Nature, Bentam Books, New York, 1984.
\bibitem{ee59} P. Ehrenfest, T. Ehrenfest, The Conceptual Foundations of the Statistical Approach in Mechanics, Cornell University Press, Ithaca, New York, 1959.
\bibitem{kryl79} N. S. Krylov, Works on the Foundation of Statistical Physics, Princeton University Press, Princeton, New York, 1979.
\bibitem{dfku04} S. Denisov, A. Filippov, J. Klafter, M. Urbakh, Phys. Rev. E 69 (2004) 042101.
\bibitem{gp71} P. Glansdorff, I. Prigogine, Thermodynamic Theory of Structure, Stability and Fluctuation, Wiley-interscience, Division of John Wiley@Son, LTD, London-New York-Sydney-Toronto, 1971.
\bibitem{zmr02} D.N. Zubarev, V.G. Morozov, G. Repke, Statistical mechnics of nonequilibrium processes. Fisoco-mathematical press, Moscow, 2002 (in Russian).
\bibitem{cw71} A.P. Cracknell, K.C. Wong, International Report IC/71/19, 1971.
\bibitem{w01} F. Willaime, La Revue de Métallurgie, (2001) 1065.
\bibitem{gt05} V. Gusev, V. Tournat, Phys. Rev. B 72 (2005) 054104.
\bibitem{c06} A. Carati, Physica A 369 (2006) 417.
\bibitem{c08} A. Carati, Physica A 387 (2008) 1491.
\bibitem{m08} L.S. Metlov, Bulletin of the Russian Academy of Sciences: Physics, 72 (2008) 1283.
\bibitem{m07} L.S. Metlov, 2007 [cond-mat] Arxiv Preprint arXiv:0711.0399.
\bibitem{vbml05} V.I. Vettregen, A.Ya. Bashkirov, G.I. Morozov, A.A. Lebedev, E.Yu. Nefed'ev, M.A. Kruchkov, Physics of Solids, 47 (2005) 1796.
\bibitem{bl93} V.S. Bobrov, M.A. Lebedkin, Physics of Solids, 35 (1993) 1881.
\bibitem{lb98} M.A. Lebedkin, L.R. Dunin-Barkovskii, Journal of Experemental and Theoretical Physics, 113 (1998) 1816.
\bibitem{pm08} V.A. Plotnikov, S.V. Makarov, Letters to Journal of Technical Physics, 34 (2008) 65.


\end{thebibliography}
\end{document}